\documentstyle[prl,amssymb,aps]{revtex}

\begin{document}

\draft
\preprint{HEP/123-qed}
\title{Spin-Landau Orbit Coupling in Units of the Flux Quantum
Observed from Zeeman Splitting of a Quantum Wire Array}
\author{J. C. Woo\footnote{E-Mail address : jcwoo@plaza.snu.ac.kr}, W. S. Kim, H. S. Ko, 
Y. A. Leem, Y. S. Kim, D. S. Kim, M. Y. Choi, and D. H. Kim}
\address{Department of Physics, Seoul National University, Seoul 151-742, Korea}
\author{T. Schmiedel\footnote{Present address : W.L. Gore and Associates, 
             Elkton, MD 21921, U.S.A.}}
\address{NHMFL, Florida State University, Tallahassee, FL 32306-4005, U.S.A.}
\date{\today}
\maketitle
\begin{abstract}
 In Zeeman spectra of a GaAs-AlGaAs quantum wire array with 
superlattice period $ l_{p}$, the zero-field shift is found to increase in steps of 
$ ( 2\pi \hbar/ e)l_{p}^{-2}$ at the 
cyclotron radius $R_{c} \approx  l_{p},  l_{p}/2$, and $l_{p}/3$.
 This shift,  caused by spin-Landau orbit coupling and quantized 
in units of the flux quantum $ 2\pi \hbar /e$, 
is a manifestation of the  gauge invariance, and the first
observation of the flux quantization in a non-superconducting material. 
The quantum interference associated with the phase difference in 
the quantum barrier scattering is also reported. 
\end{abstract}

\narrowtext

 The electronic properties of one-dimensional (1D) 
quantum structures attract particular interest. However, 
most of the studies of 1D confinement are 
limited to investigations of transport \cite{gerh89,wink89} and optical 
properties \cite{some95,ando96,some96},
 and little is known about 1D quantum confinement on the electronic spin state. 
 In this letter, we report experimental results on the Zeeman 
splitting of a quasi-1D confined electron system, namely a quantum wire (QWR) array. 
 In this system, the electronic motion is influenced by three characteristic 
physical lengths: radius $R_{c}$ of the Landau orbit,  periodic 
length $l_{p}$ of the quantum wire superlattice,  and exciton diameter. 
 The results show that the spin of the electron confined in an 1D periodic potential is 
coupled with the Landau orbital motion - this is observed as 
the zero-field splitting increases in steps for the applied field.
 The internal field produced by the Landau orbital motion gives the flux per $l_{p}^2$
quantized in units of the  flux quantum $  2\pi \hbar/ e$, which is 
characterized by $  R_{c} $ and $ l_{p}$ from the Landau gauge magnetic 
translation (LGMT).

 The Zeeman spectra are obtained from polarization dependent 
magneto-photoluminescence (PMPL) in a GaAs/Al$_{0.5}$Ga$_{0.5}$As QWR array sample. 
 The sample was grown using our Riber MBE system by migration enhanced epitaxy 
on vicinal GaAs surfaces of 1$^\circ$ off from (001) plane \cite{petr84,kim}. 
 The growth of GaAs and AlGaAs layers were switched 
every $ 1/2$ monolayer, and repeated for 30 periods. 
 The structure thus consists of GaAs QWR of 30 molecules in width 
and height with AlGaAs barriers of 30 molecules between the wires, 
providing the superlattice period $ l_{p}\approx 170$ \AA. 
 The PMPL is detected using $ \pi /2$ circular polarizer, 
while applying dc magnetic field perpendicular to the array plane. 
 The PMPL at 4.2 K is obtained in steps of l T up to 32 T with a resolution of $ \leq $ 0.05 meV. 
 The spin up and down components of the Zeeman splittings are 
obtained by reversing the field direction. 
 The TEM image of this sample and schematic diagram of the array 
structure with respect to the applied field are shown in Fig.\ \ref{fig tem} (a) and (b), respectively. 

 The PMPL peak positions $ E_{+}$  and $ E_{-}$, which correspond to 
spin up and down components respectively, are shown with $\blacktriangle$ 
and $\blacktriangledown$ symbols in Fig.\ \ref{fig pmpl}. 
 The solid line indicates the average of 
$ E_{+}$ and $ E_{-}$, which corresponds to the conventional 
diamagnetic and Landau shift. The PMPL peak at zero field is observed at a photon 
energy of 1942.5 meV. 
 A typical PMPL linewidth (FWHM) is 12 meV at low 
field and gradually decreases to 10 meV at high field. 

 The data obtained at high fields are compared with the dispersion 
of the lowest sublevel of the 2D electron system,
\begin{equation}
E_{\lambda \pm}= E_{0}+ \left(\lambda + \frac{1}{2} \right)
\frac{\hbar eB_{a}}{m^{*}}\pm \frac{1}{2}g_{eff}\mu _{B}B_{a} ,\label{elam}
\end{equation}
where + and - denote the electronic spin up and down states, 
$E_{0}$ the lowest level energy, $ \lambda (= 0,1,2,\cdots)$ the Landau level index, 
$ \omega _{c} = eB_{a}/m^{*}$ the cyclotron frequency, 
$ \mu _{B} = e\hbar /2m_{e}$, $ g_{eff}$ the effective g-factor, 
$ m^{*}$ the reduced effective mass, and $m_{e}$ the free electron mass. 
 From the slope of $ E_{\lambda ,ave} = (E_{\lambda +  }+ E_{\lambda - })/2$ 
for $ B_{a}> 10$ T, $ m^{*}= (0.145\pm 0.003)m_{e}$  is obtained. 

 Zeeman splitting is obtained from the separation of the spin up and down states, 
$ \Delta E = E_{+}-E_{-}$, and is summarized in Fig.\ \ref{fig zeeman}.
 Our results at $B_{a} \geq 4$ T show that only the electron spin in 
the ground sublevel contributes to the Zeeman splitting with little excitonic effect. 
 The microwave measurement has also reported
that the electronic spin state is hardly affected by the holes 
in an undoped quantum well (QW)  \cite{dobe88,stei83}. 
Unlike 2D cases, the Zeeman splitting of the ground-state electron can 
be fit to a single line of $ \sim g_{eff} \mu_{B} B_{a}$, 
our Zeeman data cannot be fit to a single line of $ g_{eff}\mu _{B}B_{a}$.
Using the value $ g_{eff} = -0.44$ which is the $g$-value of GaAs bulk 
and undoped GaAs QW  \cite{weis77},  
the data for $ B_{a}\geq 4$ T fit rather well with three stright lines, 
in the form
\begin{equation} 
\Delta E = g_{eff}\mu_{ B}B_{a}+\Xi. \label{delEn}
\end{equation}           
 The best fit values of $\Xi$ are 0.85 meV for 4 T $\leq B_{a}\leq $ 8 T 
and 1.70 meV for 9 T $ \leq B_{a}\leq $ 15 T. 
 For $ B_{a}\geq $ 16 T, the six data points at $B_{a}$ = 16, 17, 20, 21, 24, and 25 T, 
which are the peaks of the oscillation existing in this region, 
are used for the fit and $ \Xi $ is found to be 2.50 meV. 
(See solid lines in Fig.\ \ref{fig zeeman}.) 
 The above three regions are referred to as Regions 1, 2, and 3, respectively. 
 For these three regions, the Zeeman separation in Eq.\ (\ref{delEn}) can be rewritten as
\begin{equation} 
\Delta E_{n} = g_{eff} \mu_{ B} B_{a} + n X  \label{delEnX}
\end{equation} 
with $X = 0.85 \pm 0.02$ meV and $n = 1, 2,$ and 3.  
 The shift $ \Xi$ and thus $nX$ in Eq.\ (\ref{delEnX}) are 
mathematically equivalent to a zero-field shift in magnetic resonance, 
and the Zeeman energy $Z$ can be understood in terms of the 
internal magnetic field $B_{i}$:
\begin{equation}  
Z = \mu _{B} {\bf s } \cdot  {\bf B }_{i} ,\label{zmu}  		   
\end{equation}  
where $ {\bf s }$ is the electron spin operator.
 Since the only source for the internal field is the cyclotron motion of 
the electron, it is natural to attribute the zero-field shift to the spin-Landau orbit coupling, 
which is similar to the $ {\bf L }\cdot {\bf s }$ coupling in an atom. 
 The separation between the spin up and down states coupled with 
the internal field 
$ {\bf B }_{i} = B_{i}{\bf \hat z }$ becomes
\begin{equation}
\Delta Z =\mu _{B}({\bf s }_{\uparrow} -{\bf s }_{\downarrow})\cdot 
{\bf B }_{i}=\mu _{B}B_{i} \equiv n\mu_{B}B_{1}.\label{delgam}  
\end{equation} 
 When our experimental value of $X = 0.85 \pm 0.02$  meV is 
substituted for $ \mu_{B} B_{1}$ in Eq.\ (\ref{delgam}), it is found that
\begin{equation}  
B_{1} = 15.6 \pm 0.4{\rm T}.\label{bint}  
\end{equation}

 To explain this quantized shift, we consider a 2D electron 
in a magnetic field.  In the Landau gauge $ {\bf A } = {\bf \hat{x}}(-y)B$, 
the wavefunction at the ground-state Landau level is given by
\begin{eqnarray}  
\Psi _{k_{x}} &\equiv& \langle x, y \mid \lambda=0, k_{x} \rangle\nonumber\\
                       &= &\exp (ik_{x}x) \exp\left[-\frac{m^{*}\omega _{c}}{2\hbar }
                                   \left(y-\frac{\hbar k_{x}}{m^{*}\omega _{c}}\right)^{2}\right] \label{psik}.  
\end{eqnarray}       
 If we introduce an over-simplified 1D potential of our QWR superlattice 
$ U(y) = U_{0}\cos(2\pi y/l_{p})$, we can see that 
$ \langle k'_{x}\mid U \mid k_{x} \rangle = U_{0}\exp(-\pi /2f)\cos(k_{x}l_{p}/f)\delta _{k'_{x},k_{x}}$ 
is diagonalized with $ f \equiv eB_{a}l_{p}^{2}/(2\pi \hbar )$. 
 Comparing with the Harper equation on a 2D square lattice \cite{hofs76}, this shows that 
\begin{equation}
k_{x} = \frac{2\pi n}{l_{p}}\label{kxe} 
\end{equation}   						
is a good quantum number, which is well-defined along the $x$ direction. 
 Through the use of Eq.\ (\ref{psik}), 
the expectation value of the $z$-component of the angular momentum 
is obtained as  
\begin{eqnarray}
\langle L_{z} \rangle &=& \int \int \Psi_{k_{x}}^{*}\left[-i\hbar \left(x\frac{\partial}{\partial y}
                                        -y\frac{\partial}{\partial x} \right) \right]\Psi_{k_{x}}dxdy\nonumber\\
                               &=& \int\int \Psi_{k_{x}}^{*}
                                      \left[\hbar k_{x}y-im^{*}\omega_{c}x\left(y-\frac{\hbar                    
                                       k_{x}}{m^{*}\omega_{c}} \right) \right]\Psi_{k_{x}}dxdy\nonumber\\
                               &=&\frac{\hbar^{2}}{m^{*}\omega_{c}}k_{x}^{2}\label{lanlzran} 
\end{eqnarray}   
 A precise calculation of the effective field produced by the cyclotron motion of the 
electron in this system and coupled with the spin contributing the zero-field shift may 
require many-body consideration. 
 However, for an approximate computation, 
let us assume that the spin coupling is dominated by the field near the orbit. 
 Then, this field can be obtained via the semi-classical approach to obtain the field
\begin{equation}
B_{eff} = \frac{m^{*}v}{el_{p}} = \frac{(eB_{a}\langle L_{z} \rangle)^{1/2}}{el_{p}} 
             = \frac{2\pi \hbar}{el_{p}^{2}}n\label{Beffemv}
\end{equation} 
where Eq.\ (\ref{kxe}) and (\ref{lanlzran}) have been used.
 This show that the flux passing through the area defined by the 1D period
\begin{equation}
\Phi  = B_{eff}l_{p}^{2} = \frac{2\pi \hbar}{e}n\label{phiebeff} 
\end{equation}    			
may indeed be quantized in units of the integer flux quantum $ 2\pi \hbar /e$.

 Hofstadter has investigated the commensurability effects associated with 
the field in units of the flux 
quantum by applying the LGMT invariance to Bloch band calculations  \cite{hofs76}. 
 However, this calculation is based on a 2D periodic potential and considers the flux defined by the 
periodic spacing along both $x$ and $y$ directions. 
 In our case, even though the QWR array has only one periodic length along the $y$ direction, 
the rotational symmetry in the LGMT brings 
the addtional characteristic length along the $x$ direction.
 
 Now let us compute the flux  $B_{i}$ passing through the square $l_{p}^{2}$. 
 By substituting our experimental values of $B_{i} = B_{eff}$ and 
$l_{p} = 170$ \AA~into Eq.\ (\ref{phiebeff}), we obtain for $n = 1$ 
\begin{equation}
\Phi_{exp} = B_{1}l_{p}^{2} = (4.3 \pm 0.1) \times10^{-15} {\rm T}\cdot {\rm m}^{2}.\label{Phiext}
\end{equation}
 This value coincides with the quantity 
\begin{equation}
\Phi _{1} = \frac{2\pi \hbar} {e} = 4.13\times10^{-15} {\rm T}\cdot {\rm m}^{2}\label{Phi1e}
\end{equation}
with the error of $ \leq 5$\%. 

 The oscillatory behavior of the Zeeman shift is apparent at $ B_{a}\geq $16 T. 
(See the broken line in Fig.\ \ref{fig zeeman} which is 3 points averaging) 
 The period of this oscillation is found to be approximately $ \Delta B_{a}$ = 4.7 T 
from the peaks indicated with arrows in Fig.\ \ref{fig zeeman}. 
 This oscillatory behavior has  similarity with the quantum oscillation reported in 
magneto-transport measurements \cite{lang69,kell85} and 
the energy level calculation \cite{stei83,weis77}, 
which is characteristic of the 1D periodic potential. 
 However, the oscillatory behavior of our Zeeman splitting and 
the Landau-band conductivity are different; one is periodic with $B_{a}$ 
and the other with $1/B_{a}$.  

 This Zeeman oscillation is also a quantum oscillation caused by 1D periodicity, 
but it differs in nature. 
 Since the scattering conditions are different at points P and Q of the heterointerface, 
these two orbits have different phase factors (See Fig.\ \ref{fig schemetic}). 
 This phase difference in consequence gives the quantum interference effect to 
the net effective field (observe the differences of the solid-line path and dot line
path reflected by quantum barrier (QB) in Fig.\ \ref{fig schemetic}), and the period of the interference is 
obtained as $ l_{p}^{2}e\Delta B/\hbar  = 2\pi n$. 
 The oscillation period of $ \Delta B_{a} = 4.7$ T observed for $n = 3$ matches well with 1/3 of $B_{1}$. 
 The constructive interference would occur at the applied field corresponding to 
$ \alpha  \equiv l_{p}/R_{c} = 3$, where $ R_{c} = \sqrt{\hbar /(eB_{a})}$, 
and those separated by $ \Delta B_{a}$ from that point. 
 This is the reason why we have chosen the six peaks in Zeeman separation, 
when we fit the data for $n = 3$. 
 The reason why the transition to Region 3 occurs around $\alpha = 2.6$ 
rather than $\alpha = 3$ is not clear. 
 It may merely be due to an experimental error introduced by the fluctuation of QWR array period, 
or may be related to the interference effect. 
 The details on the quantum interference effect is out of the scope of this paper 
and will be reported elsewhere.
 
 Another feature of interest is the values of $B_{a}$ where the step-like transitions take place.
 They occur at $ \alpha = 1, 2,$ and $ \approx 3$ for the transitions to Region 1, 2 and 3, respectively. 
 The transition occuring at $n = \alpha$ means that the number of the flux quantum, 
$n$, is the maximum number of the Landau orbit which can be linearly
placed in the stripe of the length $l_{p}$. 
 This also implies that the propagation of the electron reflected by QB is modulated 
along the QWR with the period of $l_{p}$. This picture is consistent with Eq.\ (\ref{kxe}). 

 In conclusion, we have performed polarization 
dependent magneto-photoluminescence on a
GaAs/Al$_{0.5}$Ga$_{0.5}$As QWR array sample and have observed 
that the Zeeman separation increases in steps at fields 
corresponding to $\alpha = 1, 2,$ and $ \approx 3$. 
 The step-like shift is the result of  the spin coupling to 
the Landau field, the field produced by the Landau orbital motion, 
and provides the clear indication that the Landau field is manifested 
at $ B_{i} = n\Phi _{1}l_{p}^{-2}$ in units of the flux quantum $ \Phi _{1}= 2\pi \hbar /e$.  
 It is the rotational symmetry of the Landau orbit in conjunction 
with the linear translational invariance  that gives 
the additional dimensional confinement; this yields effectively
the 2D confinement condition similar to Hofstadter's. 
 Also observed along with the step-like shift in the Zeeman splitting is 
the oscillatory behavior which is apparent at $ R_{c} \lesssim l_{p}/3$. 
 The Zeeman oscillation is due to the quantum interference caused 
by the rotational phase difference of Landau orbits in QB scattering. 
 Both the Landau orbital field quantization and the quantum oscillation are 
 collective phenomena characterized by the 1D confinement and 
periodicity presented by QWR array. 
 The flux quantization observed in this well-fabricated 
QWR array provides a clear indication that a single 
electron in the Landau orbital path produces a long-range 
quantum effect in which the coherence of the 1D confinement extends over a solenoid 
definded by the superlattice period. 
 We believe that this work provides the direct and convincing evidence for
 the magnetic flux quantum in non-superconducting material. 

 The authors wish to express sincere appreciation to 
Professors J.~Ihm and D.~Heiman for helpful discussions, 
NHMFL at Florida State University for the experimental facility, 
and its staff for the priceless technical assistance. 
 This work was supported in part by BSRI 97-2421 of MOE, Korea and 
by the Korea Telecomm.

\begin{figure}
\caption{(a) TEM image of the QWR sample. 
                (b) Schematic diagram of the experimetal condition.}
\label{fig tem}
\end{figure}

\begin{figure}
\caption{The applied field dependence of peak positions of polarization 
               dependent magneto-photoluminescence spectra. 
               The symbols, $\blacktriangle$ and $\blacktriangledown$, 
               are for the spin-up and down states, respectively, 
               and the solid line shows the average between the two.}
\label{fig pmpl}
\end{figure}
        
\begin{figure}
\caption{The field dependence of Zeeman separation. 
               The solid lines are the best fit curves for the corresponding regions. 
               The arrows point to peaks in the oscillation maxima, and $ \alpha  = l_{p}/R_{c}$ is 
                marked along $x$-axis.}
\label{fig zeeman}
\end{figure}

\begin{figure}
\caption{Schematic diagram of two cyclotron paths  which have different 
                scattering conditions. It also schematically shows the path (dot line) 
                reflected at QB may present the quantum interference effect with a 
                  cyclotron path (solid line).}
\label{fig schemetic}
\end{figure}


\begin{thebibliography}{99}

\bibitem{gerh89}  R. R. Gerhardts, D. Weiss, and K. v. Klitzing, Phys. 
                  Rev. Lett. {\bf 62}, 1173 (1989).
\bibitem{wink89} R. W. Winkler, J. P. Kotthaus, and K. Ploog,
                  Phys. Rev. Lett. {\bf 62}, 1177 (1989).
\bibitem{some95} T. Someya, H. Akiyama, and H. Sakaki, Phys. Rev. Lett. 
                 {\bf 74}, 3664 (1995). 
\bibitem{ando96} H. Ando, H. Saito, A. Chavez-Pirson, H. Gotoh, and N. Kobayashi, 
                 Appl. Phys. Lett. {\bf 69}, 1512 (1996).
\bibitem{some96}  T. Someya,  H. Akiyama, and H. Sakaki,  
                 Phys. Rev. Lett. {\bf 76}, 2965 (1996).
\bibitem{petr84} P. M. Petroff, A. C. Gossard, and W. Wiegmann,
                 Appl. Phys. Lett. {\bf 45}, 620 (1984).

\bibitem{kim} Y. M. Kim, W. S. Kim, H. S. Ko, D. H. Kim, J. H. Bae, 
                 T. Schmiedel, J. Kim, J. Y. Cha, J. W. Lee, H. S. Park, S. J. Park, and J. C. Woo,
                 {\it Compound Semiconductors 1996} (IOP, Bristol, 1997), p. 303.
\bibitem{dobe88} M. Dobers, K. v. Klitzing, and G. Weimann, 
                               Phys. Rev. B {\bf 38}, 5453 (1988).
\bibitem{stei83} D. Stein, K. v. Klitzing, and G. Weimann,
                  Phys. Rev. Lett. {\bf 51}, 130 (1983).
\bibitem{weis77} C. Weisbuch and C. Hermann, Phys. Rev. B {\bf 15}, 816 (1977).
\bibitem{hofs76} D. R. Hofstadter, Phys. Rev. B {\bf 14}, 2239 (1976).
\bibitem{lang69} D. Langbein, Phys. Rev. {\bf 180}, 633 (1969).
\bibitem{kell85} M. J. Kelly, J. Phys. C {\bf 18}, 6341 (1985).


\end{thebibliography}
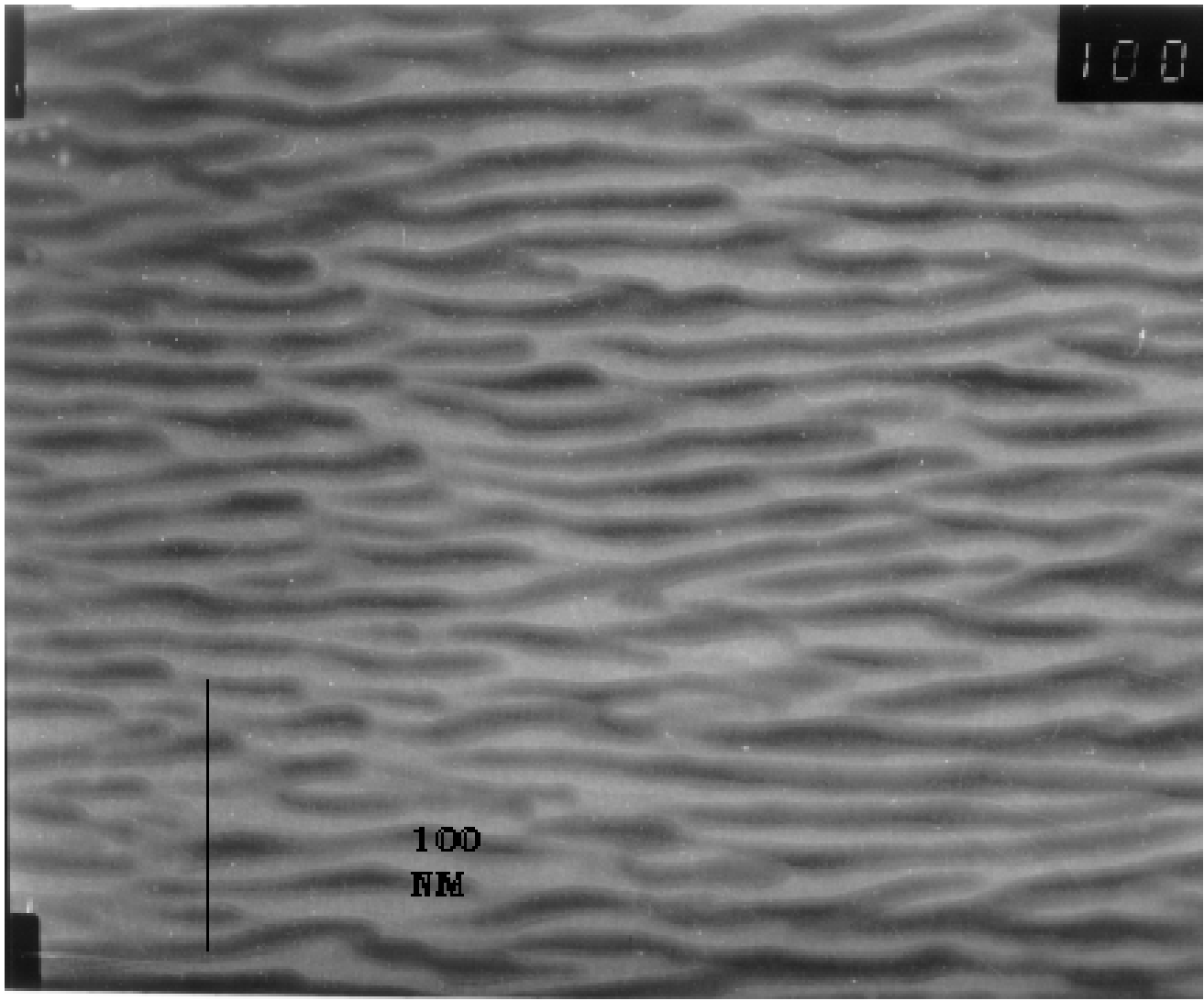
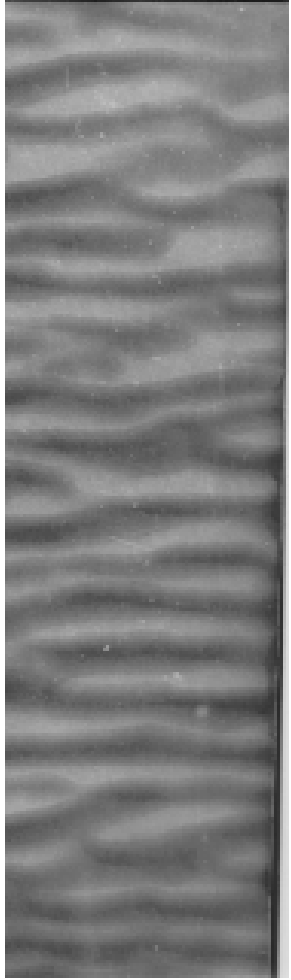
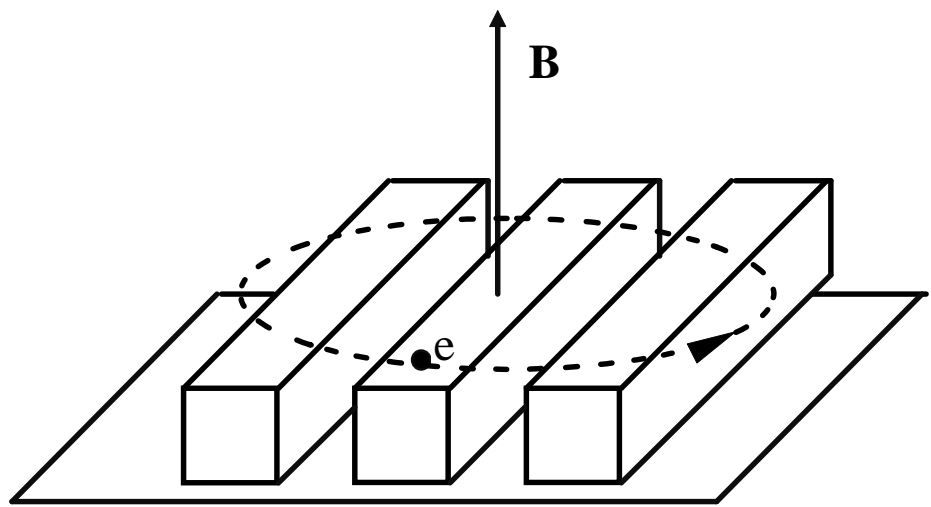
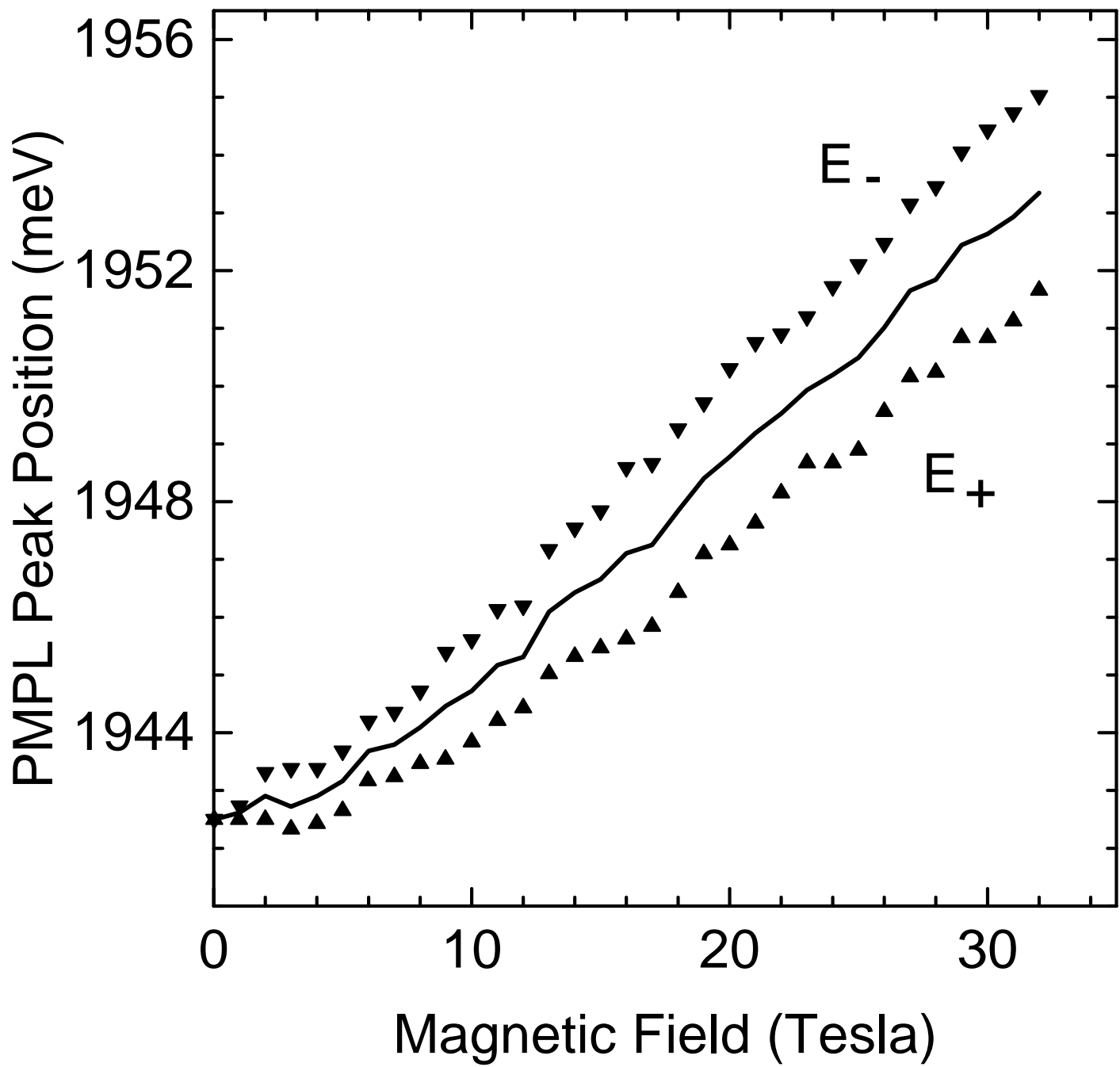
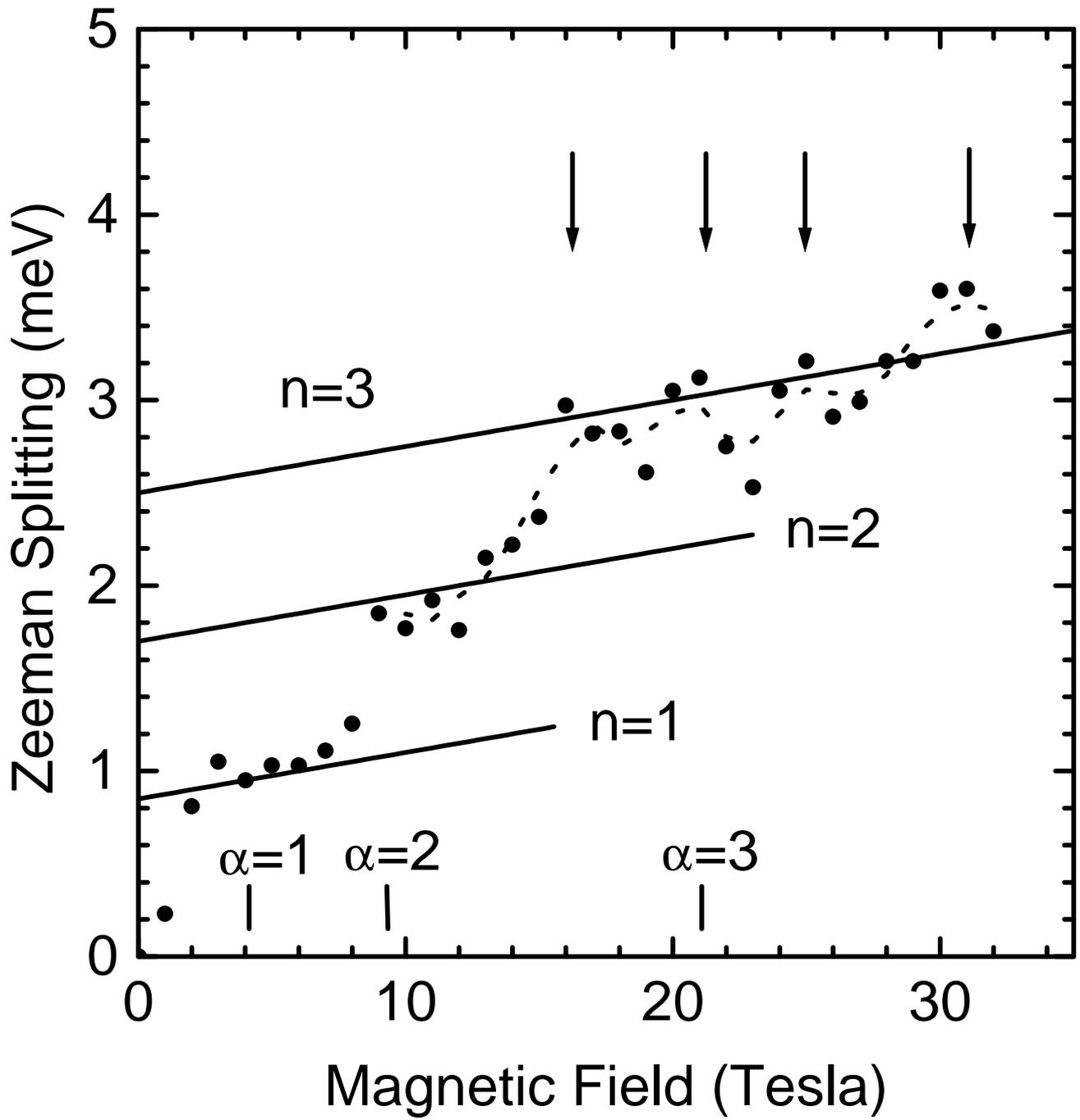
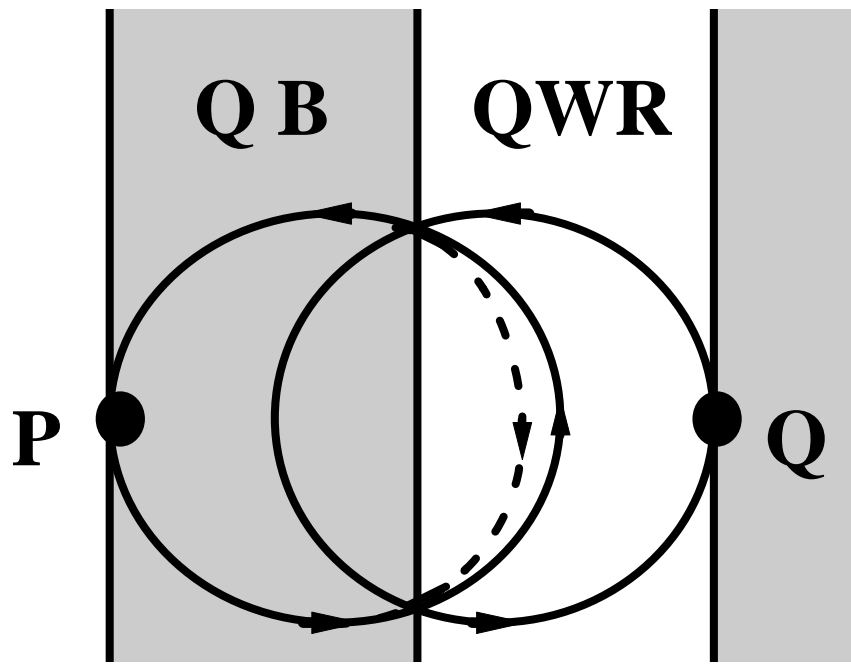
\end{document}